\documentclass[aps,prl,superscriptaddress,longbibliography,twocolumn]{revtex4-2}

\usepackage{graphicx}
\usepackage[usenames,dvipsnames]{color}
\usepackage{siunitx}
\usepackage{hyperref}
\hypersetup{colorlinks,breaklinks,
            urlcolor=[rgb]{0,0,0.64},
            linkcolor=[rgb]{0,0,0.64},
            citecolor=[rgb]{0,0,0.64},
            filecolor=[rgb]{0,0,0.64}}

\usepackage{fixltx2e}
\usepackage{upgreek,xspace}
\usepackage{amsmath}
\usepackage{amssymb}
\usepackage{amsthm}
\usepackage{mathrsfs}
\usepackage{amsbsy}
\usepackage{amstext}
\usepackage{xcolor}
\usepackage{multirow}
\usepackage{wasysym}
\usepackage{mathtools}
\usepackage{amsfonts}
\usepackage{dcolumn}
\usepackage{float}
\usepackage{bbold,bm}
\usepackage[utf8]{inputenc}

\usepackage{amssymb, amstext, amsmath}  
\usepackage{physics}                    
\usepackage{bm}                         
\usepackage{soul,xcolor}                
\usepackage{setspace}

\def\be{\begin{eqnarray}}  
\def\ee{\end{eqnarray}}

\setcitestyle{super}

\newcommand{\MITChem}{Department of Chemistry, Massachusetts Institute of Technology, Cambridge, MA 02139, USA}
\newcommand{\MITPhys}{Department of Physics, Massachusetts Institute of Technology, Cambridge, MA 02139, USA}
\newcommand{\SYSU}{State Key Lab of Optoelectronic Materials and Technologies, Guangdong Province Key Laboratory of Display Material and Technology, School of Electronics and Information Technology, Sun Yat-sen University, Guangzhou, 510275, People's Republic of China}
\newcommand{\UCBChem}{Department of Chemistry, University of California, Berkeley, CA 94720, USA}
\newcommand{\HarvardPhys}{Department of Physics, Harvard University, Cambridge, MA 02138, USA}
\newcommand{\Rutgers}{Department of Physics and Astronomy, Center for Materials Theory, Rutgers University, Piscataway, NJ 08854, USA}
\newcommand{\MITEECS}{Department of Electrical Engineering and Computer Science, Massachusetts Institute of Technology, Cambridge, MA 02139, USA}
\newcommand{\MITCMS}{Center for Materials Science \& Engineering, Massachusetts Institute of Technology, Cambridge, MA 02139, USA}
\newcommand{\NIMS}{International Center for Materials Nanoarchitectonics, National Institute for Materials Science,  1-1 Namiki, Tsukuba 305-0044, Japan}
\newcommand{\RCFM}{Research Center for Functional Materials, National Institute for Materials Science, 1-1 Namiki, Tsukuba 305-0044, Japan}
\newcommand{\UTAustin}{Department of Physics, Center for Complex Quantum System, The University of Texas at Austin, Austin, TX 78712, USA}
\newcommand{\EqualContribute}{These authors contributed equally to this work: J. Shi, Y.-Q. Bie, and A. Zong.}

\begin{document}

\title{Intrinsic 1\textit{T}$'$ phase induced in atomically thin 2\textit{H}-MoTe$_\text{2}$ by a single terahertz pulse}

\author{Jiaojian Shi}
\thanks{\EqualContribute}
\affiliation{\MITChem}

\author{Ya-Qing Bie}
\thanks{\EqualContribute}
\email{bieyq@mail.sysu.edu.cn}
\affiliation{\MITPhys}
\affiliation{\SYSU}

\author{Alfred Zong}
\thanks{\EqualContribute}
\affiliation{\UCBChem}
\affiliation{\MITPhys}

\author{Shiang Fang}
\affiliation{\HarvardPhys}
\affiliation{\Rutgers}

\author{Wei Chen}
\affiliation{\HarvardPhys}

\author{Jinchi Han}
\affiliation{\MITEECS}

\author{Zhaolong Cao}
\affiliation{\SYSU}

\author{Yong Zhang}
\affiliation{\MITCMS}

\author{Takashi Taniguchi}
\affiliation{\NIMS}

\author{Kenji Watanabe}
\affiliation{\RCFM}

\author{Vladimir Bulovi\'c}
\affiliation{\MITEECS}

\author{Efthimios Kaxiras}
\affiliation{\HarvardPhys}

\author{Edoardo Baldini}
\affiliation{\MITPhys}
\affiliation{\UTAustin}

\author{Pablo Jarillo-Herrero}
\email{pjarillo@mit.edu}
\affiliation{\MITPhys}

\author{Keith A. Nelson}
\email{kanelson@mit.edu}
\affiliation{\MITChem}

\date{\today}

\begin{abstract}
Polymorphic transitions in layered transition metal dichalcogenides provide an excellent platform for discovering exotic phenomena associated with metastable states, ranging from topological phase transitions \cite{Sie2019} to enhanced superconductivity \cite{Dong2021}. In particular, the transition from 2\textit{H} to 1$T'$-MoTe$_2$, which was thought to be induced by high-energy photon irradiation \cite{Cho2015,Song2018,Tan2018} among many other means \cite{wang2017,Zhang2019,qi2016,hou2019strain},  has been intensely studied for its technological relevance in nanoscale transistors \cite{Cho2015}. Despite the remarkable electrical performance arising from this 2$H$-to-1$T'$ transition, it remains controversial whether a crystalline 1$T'$ phase is produced because optical signatures of this putative transition are found to be associated with the formation of elemental Te clusters instead \cite{Chen2017,manjon2021anomalous}. Here, we demonstrate the creation of an intrinsic 1$T'$ lattice after irradiating a mono- or few-layer 2$H$-MoTe$_2$ with a single field-enhanced terahertz pulse, whose low photon energy limits possible structural damage by optical pulses. To visualize the temporal evolution of this irreversible transition, we further develop a single-shot terahertz pump-second harmonic probe technique, and we find that the transition out of the 2$H$ phase occurs within 10~ns after photoexcitation. Our results not only resolve the long-standing debate over the light-induced polymorphic transition in MoTe$_2$, they also highlight the unique capability of strong-field terahertz pulses in manipulating the structure of quantum materials.
\end{abstract}

\maketitle

\begin{figure*}[htb!]
	\includegraphics[width=0.72\textwidth]{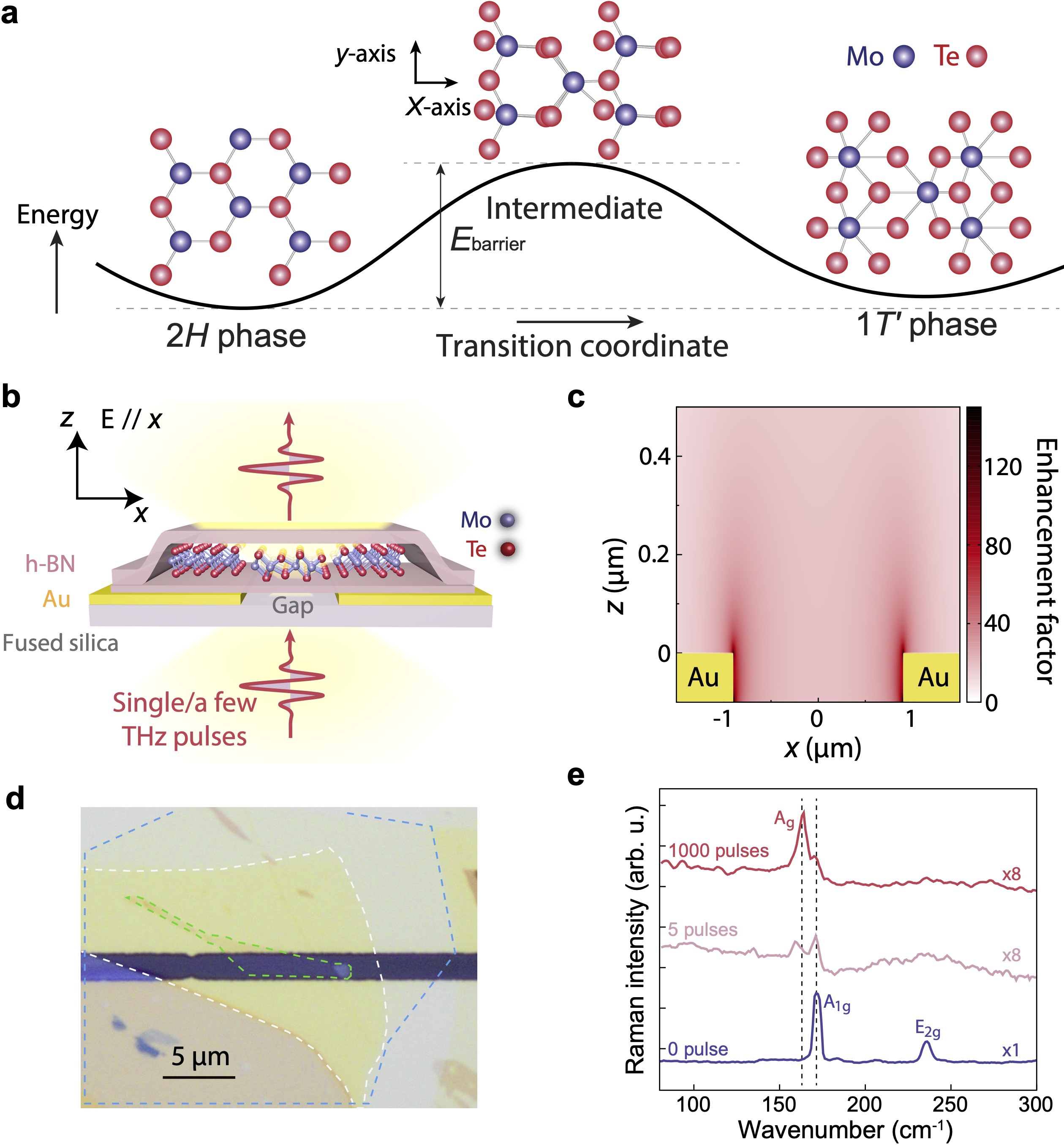}
	\caption{\textbf{Polymorphic transition in MoTe$_\text{2}$ induced by high-field THz pulses.} \textbf{a},~Schematic energy landscape and top view of lattice structures of 2$H$-MoTe$_2$ (left), the intermediate state (middle), and 1$T'$-MoTe$_2$ (right). \textbf{b},~Cross-sectional schematic illustration of a 2$H$-MoTe$_2$ crystal spanning an insulating gap between deposited gold strips, which serve as a THz field enhancement structure. The monolayer 2$H$-MoTe$_2$ crystal is encapsulated between top and bottom h-BN. THz pulses are incident from the side of the fused silica substrate. \textbf{c},~Field strength calculation of the THz enhancement structure. Numerical simulation results showing THz field enhancement by a factor of 20--50 in significant regions in and near the gap between the gold strips. The \emph{enhancement factor} is defined as the ratio between the actual and incident electric field. \textbf{d},~Optical micrograph of a monolayer sample, including a 15-nm-thick top h-BN (white dashed line), a monolayer 2$H$-MoTe$_2$ (green dashed line), and a 5-nm-thick bottom h-BN (blue dashed line), all spanning a 1.8-$\upmu$m insulating gap (dark horizontal line) between the top and bottom gold strips. \textbf{e},~Raman spectra of monolayer MoTe$_2$ after successive THz pulse irradiation with free-space field amplitudes of 270~kV/cm. The $A_g$ mode of the 1$T'$ phase sets in after 5 THz pulses and dominates upon further irradiation. By contrast, the $A_{1g}$ and $E_{2g}$ modes of the 2$H$ phase are strongly reduced. The Raman measurements were conducted with a 1-$\upmu$m-diameter laser spot and were therefore averaged over regions that had been subjected to differently enhanced THz field strengths.}
\label{fig:1}
\end{figure*}

Tailored, ultrashort pulses of light can be used to manipulate metastable states in functional materials, such as triggering insulator-to-metal transitions \cite{Liu2012}, unveiling hidden states \cite{Stojchevska2014}, or even inducing long-lasting superconductivity way above the equilibrium critical temperature \cite{Budden2021}. These photoinduced states have been intensely studied not only because they yield profound insights into fundamental light-matter interactions in correlated systems but also because they bring new opportunities to the field of laser fabrication and micro-machining, going beyond traditional laser cutting \cite{Niziev_1999}, burning \cite{BUCHALLA2007586}, or stereolithography technologies \cite{popov2004laser}. From this application perspective, the metastable 1$T'$ phase of MoTe$_2$ induced in a semiconducting 2$H$ polymorph is considered an outstanding candidate for tackling the issue of high contact resistance for two-dimensional electrical device \cite{zhang2019low}. In this polymorphic transition, the hexagonal unit cell of 2$H$-MoTe$_2$ in its ground state is transformed into the metastable 1$T'$ phase after going through an intermediate structure (Fig.~\ref{fig:1}a), a process that was believed to be readily instigated by optical irradiation \cite{Kolobov2016,Peng2020}. Compared to other methods such as ionic gating \cite{wang2017,wang2021direct} and strain application \cite{hou2019strain}, the focused laser spot down to the sub-micrometer regime makes an optically driven transition especially appealing for precision fabrication of small devices.

Despite the high technological impact of this polymorphic transition in MoTe$_2$, recent studies have cast doubt on the existence of the purported 1$T'$ structure upon optical photon illumination. In particular, the main evidence for the 1$T'$ polymorph was the $A_g$ phonon peaks around 120~cm$^{-1}$ and 140~cm$^{-1}$ in Raman spectroscopy \cite{Chen2017,manjon2021anomalous} (Figs.~S1 and S2), which have been shown to originate from nano- or micro-sized Te clusters. Given the high energy barrier $E_\text{barrier}$ of 0.88~eV per MoTe$_2$ formula unit \cite{duerloo2014structural} as well as the extremely large Te displacement of more than 1~{\AA} required for the transition \cite{Keum2015} (Fig.~\ref{fig:1}a), it remains unclear whether light can initiate the polymorphic change after all. 

In this work, instead of using visible light, we report the successful transition from 2$H$ to 1$T'$-MoTe$_2$ using single-cycle, field-enhanced terahertz (THz) pulses. The transition can be induced with as few as a single THz pulse on atomically thin crystals of 2$H$-MoTe$_2$. Using Raman spectroscopy, we show that the induced phase reflects an intrinsic 1$T'$ structure \cite{wang2017,wang2021direct}, which is free from any spurious features associated with Te clusters. The validity of the metastable 1$T'$ phase is further confirmed by selected area electron diffraction, which demonstrates a macroscopic structural transition. Using single-shot, time-resolved nonlinear optical spectroscopy, we also unveil the different stages of this irreversible phase transition. These findings provide a new route of structural manipulation using high-field THz pulses, which crucially avoids large-scale defect generation that often accompanies intense optical excitation \cite{Cho2015}.

\newpage
\begin{figure*}[htb!]
	\includegraphics[width=0.9\textwidth]{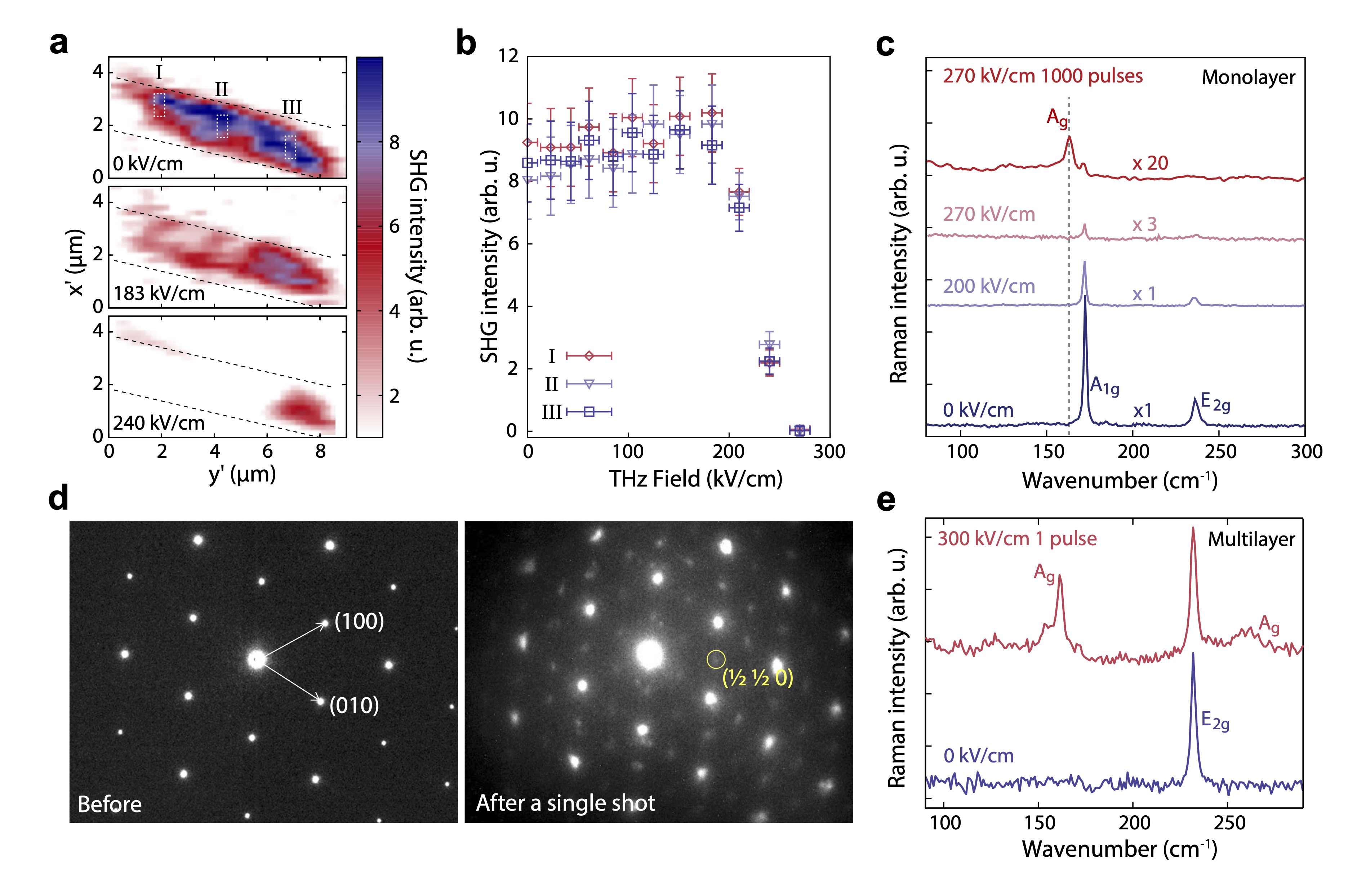}
	\caption{\textbf{THz field dependence of the phase transition in monolayer and multilayer MoTe$_\text{2}$.} \textbf{a},~SHG images of monolayer MoTe$_2$ before THz irradiation (0~kV/cm), after irradiation with one THz pulse at 183~kV/cm free-space field amplitude, and after irradiation with a second THz pulse at 240~kV/cm. The dashed lines indicate the edges of gold strips in the field enhancement structure. Different areas of monolayer MoTe$_2$ in the gap are labeled by I, II and III. The primes in $x'$ and $y'$ of the axes are added to differentiate the $x$-$y$ coordinates in Fig.~\ref{fig:1} due to the rotated field of view in SHG microscopy. \textbf{b},~SHG from different spots (I--III) in \textbf{a} measured after irradiation of the sample by single THz pulses with successively increasing field strength. The SHG signal starts to decrease at a free-space field strength of 200~kV/cm and disappears above 240~kV/cm. \textbf{c},~Raman spectra of monolayer MoTe$_2$ samples prior to THz irradiation (blue curve), after THz irradiation with a single pulse at 200~kV/cm (violet curve), another single pulse at 270~kV/cm (pink curve), and 1000 pulses at 270~kV/cm (red curve). The Raman modes of the 2$H$ phase start to decrease after a single THz pulse above 200~kV/cm. The monolayer MoTe$_2$ sample shows the $A_g$ mode at 163.5~cm$^{-1}$ after irradiation with 1000~pulses at 270~kV/cm. \textbf{d},~Electron diffraction pattern of a multilayer ($\sim10$~layers) MoTe$_2$ before and after a single THz pulse irradiation at 300~kV/cm, showing the emergence of superstructure peaks (yellow circle) that are characteristic of the $1T'$ phase. \textbf{e},~Raman spectrum of the multilayer MoTe$_2$ before and after a single THz pulse irradiation at 300~kV/cm, showing the emergence of new Raman peaks that are characteristic of the induced 1$T'$ phase.}
\label{fig:2}
\end{figure*}

\newpage
\begin{figure*}[htb!]
	\includegraphics[width=0.9\textwidth]{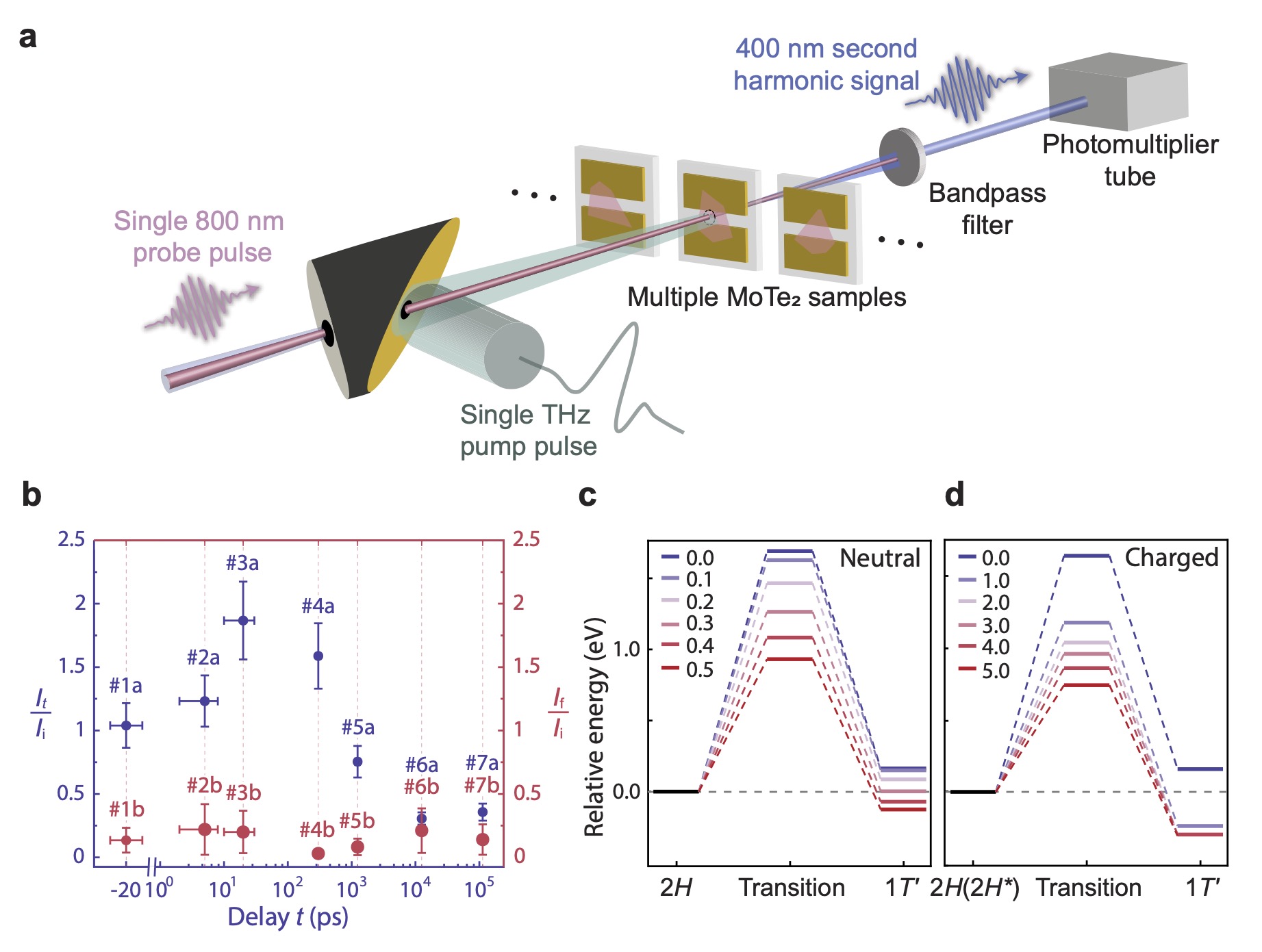}
	\caption{\textbf{Dynamics and driving mechanism of the THz-field-induced phase transition.} \textbf{a},~Schematic illustration of the setup for single-shot SHG probe with a THz excitation pulse. Single-shot measurements were conducted by using a fresh sample in each shot. \textbf{b},~SHG intensities measured from trilayer MoTe$_2$ samples at different delay times $t$ from several picoseconds to hundreds of nanoseconds (blue dots) as well as around 1 minute (red dots). The data points on the same dashed line show the SHG intensity measured from the same flake shortly after ($I_t$) and 1~min after the THz excitation pulse ($I_f$), normalized by the SHG signal intensity prior to THz excitation ($I_i$). Due to the destructive nature of single-shot SHG, $I_i$, $I_t$, and $I_f$ were measured at three different spots on the same sample. For this reason, we use \#1a, \#1b, etc. to emphasize that these data points were taken at different spots (a,~b) of the same flake (\#1). The error analysis is provided in Supplementary Note~8. \textbf{c},~The energy potential as a function of transition coordinate with nonequilibrium carrier distribution (charge neutral). Nonequilibrium distribution of carriers is qualitatively described by the Fermi-smearing method. The free-energy barrier decreases from 1.66~eV to 0.91~eV as the Fermi-smearing width increases from 0 to 0.5~eV. The figure legend has a unit of eV. \textbf{d},~The energy potential as a function of transition coordinate upon charge dopings. As the added charge density increases to 1.0~$e/$MoTe$_2$, which corresponds to $9\times10^{14}$~cm$^{-2}$, the activation energy decreases from 1.66~eV to 1.19~eV. The figure legend has a unit of $e/$MoTe$_2$. The atomic structures of transition states are shown in Figs.~S16 and S17.}
\label{fig:3}
\end{figure*}

In our experiments, we positioned a flake of 2$H$-MoTe$_2$ encapsulated by hexagonal-boron nitride (h-BN) on top of two parallel gold strips separated by a gap (Fig.~\ref{fig:1}b). The h-BN layers were used to isolate monolayers and bilayer MoTe$_2$ samples from field-induced emission in the gold layers \cite{Dean2010} and to prevent their direct exposure to air. Figure~\ref{fig:1}d shows the optical image of a monolayer sample prepared using a dry transfer method \cite{Wang2013}. Upon illumination of this structure with a free-space THz pulse, the field strength can be enhanced by more than 100~times at certain hot spots and about 20~times at the gap center, as shown in Fig.~\ref{fig:1}c.  Since the peak electric field of the THz pulse in free space reached $\sim 300$~kV/cm at the focus, the electrical field amplitude can be more than 10~MV/cm in significant regions of the sample \cite{Pein2017} (see Methods, Supplementary Note~1, and Fig.~\ref{fig:1}c for detailed information on THz generation).  After irradiating this structure with a free-space THz field of 270~kV/cm, we probed lattice changes via spontaneous Raman scattering. Figure~\ref{fig:1}e shows the Raman spectrum before the THz illumination. It consists of the $A_{1g}$ phonon at 171.5~cm$^{-1}$ and the $E_{2g}$ phonon at 236~cm$^{-1}$, in agreement with previous studies of the 2$H$ phase \cite{Keum2015}. After excitation with five THz pulses, both modes of the 2$H$ phase disappear and a peak at 163.3~cm$^{-1}$ emerges (see Fig.~S3 for additional peak width analysis). This feature signifies a change of the crystalline symmetry, and it is considered a specific fingerprint of the 1$T'$ phase of MoTe$_2$ \cite{Keum2015,wang2017}. The new phonon peak becomes more prominent after further exposure to THz pulses (Fig.~\ref{fig:1}e), indicating the growth of a larger 1$T'$ sample area. Importantly, the spectrum does not show any localized modes due to damage-related Te clusters \cite{Chen2017} even though the resultant $1T'$ phase is not homogeneously produced in the flake suspended between the gap of the gold strips (Figs.~S4 and S5). These observations were reproduced in multiple samples, suggesting that the action of a strong THz field on our monolayer MoTe$_2$ is to induce a phase transition from the 2$H$ to the 1$T'$ polymorph.

To further investigate this structural transformation, we conducted experiments on a series of monolayer 2$H$-MoTe$_2$ samples. Since the induced 1$T'$ phase is long-lived, each measurement required a fresh specimen that was carefully positioned over the gap between the gold strips, as indicated in Fig.~\ref{fig:1}b. First, we studied how the 2$H$ phase responded to a sequence of THz pulses with a gradually increasing field strength. To probe the phase transition, we tracked the changes in crystalline symmetry via optical second harmonic generation (SHG), a sensitive probe of inversion symmetry breaking that can distinguish between the non-centrosymmetric 2$H$ phase and the centrosymmetric 1$T'$ phase in samples with an odd number of layers \cite{Yamamoto2014,Beams2016} (see Methods). Figure~\ref{fig:2}a shows three representative real-space images of the SHG intensity before and after THz-field exposures with free-space amplitudes of 183~kV/cm and 240~kV/cm; additional SHG intensity images are shown in Fig.~S6. A plot of the SHG intensities from three sample locations within the field-enhancement gap (labeled I, II, III in Fig.~\ref{fig:2}a) is presented in Fig.~\ref{fig:2}b. We observe that the SHG signal drops when the field strength exceeds 183~kV/cm, disappearing completely above 270~kV/cm. As the field enhancement factor generated by the gold strips is approximately 20, the threshold needed to quench the SHG intensity is estimated to be around 4.0~MV/cm. We complemented these measurements by recording Raman spectra as a function of the THz field strength. Figure~\ref{fig:2}c shows the spectra acquired after irradiation at 200~kV/cm and 270~kV/cm. Consistent with the picture supported by the SHG results in Fig.~\ref{fig:2}b, a single-pulse irradiation first leads to a drop in the intensity of the 2$H$ phonons with a free-space field strength of $200$~kV/cm, and stronger THz pulses at 270~kV/cm subsequently cause a further suppression, leading to a nearly complete extinction of the peak after 1000 pulses. Simultaneously, the Raman mode of the 1$T'$ phase grows, eventually dominating the spectrum.

The THz-induced transition is not restricted to monolayer samples, and we observed similar phenomenology in bilayer and multilayer ($\sim10$ layers) 2$H$-MoTe$_2$ (see Figs.~S5c,d and S7c,d). Figures~\ref{fig:2}d and \ref{fig:2}e show the results obtained on a multi-layer 2$H$-MoTe$_2$, presenting the [001] zone-axis electron diffraction pattern and Raman spectrum before and after irradiation with a single THz pulse at 300~kV/cm. We find that the diffraction pattern of the unexcited crystal exhibits the six-fold symmetry expected for the 2$H$ phase. Consistent with this observation, the Raman spectrum displays only an $E_{2g}$ phonon mode of bulk 2$H$-MoTe$_2$ \cite{Ruppert2014}. After THz irradiation, new peaks emerge in the diffraction pattern taken along the [001] zone axis of the new unit cell, revealing a cell-doubling superstructure \cite{Wang2014,Gao2015,Heising1999,Meng2019}. This diffraction change is accompanied by the appearance of $B_{g}$ and $A_{g}$ phonons in Raman spectroscopy \cite{Keum2015,wang2017}. Both observables are characteristic signatures of the 1$T'$ lattice, offering additional validation for the creation of an intrinsic 1$T'$ phase (see Supplementary Notes~2--3 and Fig.~S8).

To gain microscopic insights into the THz-driven transition, we next examine the photo-induced dynamics to trace out the temporal evolution of the crystalline lattice. This investigation cannot be accomplished by conventional time-resolved spectroscopy methods because their stroboscopic approach precludes the study of irreversible processes. We therefore developed a specialized apparatus for THz pump-SHG probe single-shot spectroscopy, which is well suited for probing the formation dynamics of the long-lived metastable 1$T'$ state. A schematic illustration of our experiment is presented in Fig.~\ref{fig:3}a; details of the setup are described in Methods. In our measurements, we irradiated trilayer samples with a single THz pulse, whose field strength was adjusted to exceed the phase transition threshold. We subsequently recorded the SHG signal at different pump-probe delays. Since the THz field was above the threshold for generating the 1$T'$ phase, the experiment was challenging because each THz shot with its predetermined delay time needed to be conducted on a fresh piece of sample. An added difficulty is the extremely weak single-shot SHG signal from a few-layer sample; using a photomultiplier tube, we obtained measurable second harmonic intensity under an incident 800-nm pulse fluence of $\sim20$~mJ/cm$^2$, which is beyond the optical damage threshold. Here, the collection of single-shot SHG signals relies on the concept of ``probe before destruction'' \cite{Chapman2011,Tenboer2014}, which is well applicable to MoTe$_2$ (see Supplementary Note~8). The small beam spot for SHG ($1~\upmu$m) relative to the sample size ($\geq10~\upmu$m) and good spatial uniformity of the trilayer crystals (Fig.~S12) allow a comparison among three locally destructive single-shot SHG measurements at different sample locations: one prior to THz excitation ($I_i$), the second at the selected delay time following THz excitation ($I_t$), and the third at a long delay time at approximately 1~min ($I_f$). 

Figure~\ref{fig:3}b shows the time evolution of the SHG signal relative to the initial signal following THz excitation ($I_t/I_i$, blue dots). The response first increases, reaching about twice the initial value at 20~ps. It then remains higher than $I_i$ for several hundred picoseconds, before dropping to $\sim0.2 I_i$ at 12.5~ns. Finally, at long pump-probe delays, the SHG signal disappears completely, indicating the stabilization of a metastable phase. The metastability is further confirmed by the value of the final signals after 1~min of photoexcitation, $I_f/I_i$, shown as the red dots in Fig.~\ref{fig:3}b. We were able to reproduce these complex dynamics on different samples, with the initial enhancement of the SHG occurring even below the phase transition threshold (Fig.~S9). From our single-shot spectroscopy data, we identify a characteristic timescale on the order of 10~ns that is needed to completely switch MoTe$_2$ out of the 2$H$ polymorph. This timescale is comparable to the reported carrier recombination lifetimes in few-layer MoTe$_2$ \cite{Chen2018} (see Supplementary Note~4). We attribute the increase in SHG observed at shorter times to a possible nucleation of a transient structure with a lower crystalline symmetry than 2$H$, for example, the distorted trigonal prismatic 2$H^*$ phase \cite{Kolobov2016,Zhang2019}; this scenario at short delay times requires additional characterization by single-shot diffraction experiments, which are beyond the scope of our present study.

To rationalize our findings, we theoretically explored different effects that can account for the 2$H$-to-1$T'$ transition. Based on the nanosecond phase transition timescale, a natural hypothesis involves a THz field-driven carrier excitation mechanism. Microscopically, an intense electric field at THz frequency can generate high carrier densities through mechanisms such as Poole-Frenkel ionization, which liberate carriers and accelerate them to multi-eV energies (see Supplementary Note~5). These processes lead to impact ionization, liberating more carriers and evolving as a cascade event \cite{Fan2013}. In our experiments, carriers can either remain as quasi-free in the bands or localize at the substrate/h-BN interfacial states. In the latter case, they lead to a local charging of the MoTe$_2$ layers\cite{Ju2014}. To account for both scenarios, we investigated the effects of a nonequilibrium neutral carrier redistribution and those of charge doping on the energy landscape of MoTe$_2$. 

The results of our calculations are shown in Fig.~\ref{fig:3}c,d (see Methods and Supplementary Note~6 for details of the first-principles calculations). In equilibrium, the metastable 1$T'$ phase lies about 0.1~eV higher in energy than the 2$H$ phase (blue horizontal lines); here, energy values are quoted per two formula units of MoTe$_2$ due to unit cell doubling in this transition. The 2$H$-to-1$T'$ activation barrier and the reverse activation barrier are 1.66~eV and 1.56~eV, respectively, thus protecting the 2$H$ and 1$T'$ phases against thermal fluctuations. Upon excitation of a neutral electron-hole redistribution (Fig.~\ref{fig:3}c) or charge doping (Fig.~\ref{fig:3}d), two effects occur. First, the activation barrier lowers substantially (violet-to-red horizontal lines) and second, the 1$T'$ energy decreases, both effects favoring the occurrence of the 2$H$-to-1$T'$ transition. Further analysis shows that the neutral carrier redistribution takes the leading role in this process compared to the charge doping scenario. This is because the density of interfacial traps needed for charge doping is three orders of magnitude smaller than the effective charge densities required to stimulate the phase transition, as explained in more detail in Supplementary Note~5. 

Our joint experimental and theoretical studies establish intense THz fields as a viable route to steer the polymorphic transition in atomically thin MoTe$_2$. Crucially, this transition cannot be induced by photons at higher energies \cite{Cho2015,Song2018,Tan2018,Krishnamoorthy2019}, as evidenced by our experiments performed with mid-infrared and near-infrared pulses centered around 225~meV and 1.55~eV, respectively (Figs.~S1 and S2). This large structural distortion leads to a significant electronic structure reconstruction involving band inversion around the $\Gamma$ point, driving a topologically trivial 2$H$ phase of few-layer MoTe$_2$ into a quantum spin Hall insulator state  in the 1$T'$ phase \cite{Qian2014} (see Supplementary Note~7). The capability demonstrated in this work hence opens the tantalizing prospect in the search for novel phases of matter with non-trivial band topology. Our findings further highlight the unique capability of high-field THz pulses in shaping material properties in a complex, multi-phase energy landscape.

\providecommand{\noopsort}[1]{}\providecommand{\singleletter}[1]{#1}%

\newpage
\section{Methods}

\noindent \textbf{Fabrication of THz field-enhancement structure on fused silica substrate}\\
The fabrication of the metal microslit array was based on a standard photolithography and lift-off process. Image reversal photoresist AZ5214 was spin-coated on a fused silica substrate at 3000~rpm for 30~s, soft baked at 110$^{\circ}$C for 50~s on a hotplate, UV exposed by a maskless aligner MLA~150 with a dose of 24~mJ/cm$^2$, and post-exposure baked at 120$^{\circ}$C for 2~min followed by flood exposure and development in AZ422. A thin film of 2-nm Cr was deposited onto the substrate as an adhesion layer by thermal evaporation followed by a 98-nm-thick Au thin film. The sample was soaked in acetone and PG remover for lift-off to complete fabrication of the field enhancement structure.\\

\noindent \textbf{Layered MoTe$_\text{2}$ integration with the field enhancement structure}\\
The monolayer, few-layer MoTe$_2$, and layered h-BN were exfoliated on SiO$_2$/Si substrate from bulk MoTe$_2$ (HQ graphene) or h-BN crystals. Monolayer and bilayer MoTe$_2$ were identified by optical contrast and Raman spectroscopy. The layered materials were picked up by a transfer slide composed of a stack of glass, a polydimethylsiloxane (PDMS) film and a polycarbonate (PC) film, as described in ref.~\cite{Zomer2014}. The resulting stacks of top h-BN layer, MoTe$_2$ monolayer or bilayer, and bottom h-BN layer were then placed on top of the Au gap with the help of a transfer setup under an optical microscope \cite{Dean2010}. h-BN layers were not used for MoTe$_2$ trilayer and multi-layer samples because they will be totally damaged by the optical pulse for SHG if it is encapsulated by h-BN.\\

\noindent \textbf{High-field THz pulse generation}\\
High-field THz pulses were generated in a Mg:LiNbO$_3$ crystal by tilting the optical pulse front of an 800-nm pump pulse to achieve phase matching \cite{Hebling2008}. By using a three-parabolic-mirror THz imaging system, the image of the THz beam spot on the sample was focused to its diffraction limit of around 500~$\upmu$m in diameter. The incident THz pulse temporal profile was measured in the time domain using electro-optic sampling with a 100-$\upmu$m-thick 110-oriented GaP crystal. When pumping with the 4~W output from an amplified Ti:sapphire laser system (repetition rate 1~kHz, central wavelength 800~nm, pulse duration 100~fs), the peak electric field of the THz pulses reached $\sim 300$~kV/cm at the focus, with a spectrum centered at around 0.5~THz (see Figs.~S10 and S11). The repetition rate of the laser output was down-counted from 1~kHz to 125~Hz by three successive phase-locked choppers, providing sufficient temporal separation between pulses for a mechanical shutter to isolate single THz pulses.\\

\noindent \textbf{Spontaneous Raman scattering measurements}\\
Spontaneous Raman scattering was performed on samples before and after THz irradiation using a commercial Raman spectrometer (Horiba LabRAM) with a HeNe laser ($\lambda=632.8$~nm). The laser beam was focused on the samples by a 100$\times$ objective into a spot with a diameter of approximately 1~$\upmu$m.\\

\noindent \textbf{Transmission electron diffraction microscopy}\\
To prepare the samples for transmission electron microscopy (TEM) characterization before and after THz excitation, we designed a special TEM grid with a THz field enhancement structure. The enhancement pattern made of a 50-nm-thick gold film is deposited on a silicon nitride TEM window. The enhancement structure is a 300-$\upmu$m-long, 2-$\upmu$m-wide air gap within the gold film, as shown in Fig.~S8. We transferred a multilayer MoTe$_2$ flake on top of the gap using a PDMS dry-transfer method \cite{Bie2021} and measured the diffraction pattern before and after the single THz pulse excitation. The transmission electron diffraction measurement was performed using a transmission electron microscope (FEI Tecnai Multipurpose Digital TEM) operated at 60~keV at room temperature.\\

\noindent \textbf{SHG mapping}\\
As shown in Fig.~S11, the SHG fundamental pulses were provided by a mode-locked Ti:sapphire oscillator centered at 800~nm. The laser pulse duration was around 35~fs at an 80~MHz repetition rate. The pulse was linearly polarized by an achromatic polarizer (400--800~nm) and the polarization at the sample was controlled by an achromatic half-wave plate in a motorized rotational stage. SHG signals from the sample were collected by the same objective for focusing the fundamental light, and they transmitted through the same half-wave plate and polarizer, which ensured that the SHG components detected were parallel to the polarization of the fundamental field. A photomultiplier tube (PMT, Hamamatsu Photonics H10721) was used to analyze the SHG signal. In the fast mapping mode, the laser beam on the sample was scanned using two-axis Galvo mirrors (Thorlabs, GVS412) to acquire \textit{in situ} SHG images. We also use the SHG mapping method to look for trilayer samples which have uniform response as shown in Fig.~S12 for the single-shot THz pump-SHG probe measurements.\\

\noindent \textbf{Single-shot THz pump-SHG probe microscopy}\\
The THz pump arm was combined with SHG pulses from a second, temporally synchronized 12~W amplified Ti:sapphire laser (repetition rate 1~kHz, central wavelength 800~nm, pulse duration 35~fs). SHG light was focused to a near-diffraction-limited size (1~$\upmu$m) at the sample with a 50$\times$ objective. The SHG light was collected by the same objective and detected by a PMT with a confocal microscope to selectively probe the area at the focus with single-optical-pulse irradiation. The power level of the SHG excitation pulse was well above the damage threshold of MoTe$_2$, but the ultrafast nature of the pulse enabled us to obtain a reliable SHG signal before the sample was damaged. The temporal overlap of the counter-propagating THz field and SHG optical pulse was determined by THz field-induced second harmonic signal from a 30-$\upmu$m-thick LiNbO$_3$ slab. All of the optical measurements were conducted on a single-shot basis under ambient conditions. More details are given in Supplementary Note~8 and Figs.~S12d, S13, and S14.\\

\noindent \textbf{First-principles calculations}\\
We performed density functional theory (DFT) calculations using the VASP code \cite{Kresse1996,Kresse1996b} within the general gradient approximation (GGA) - Perdew-Burke-Ernzerhof (PBE) exchange-correlation functional \cite{Perdew1996}. The kinetic energy cutoff was 350~eV for the plane-wave basis sets. The lattice constants of 2$H$ and 1$T'$ phases in the neutral state were obtained via structural optimization with a convergence threshold of force on each atom of 0.01~eV/{\AA}, with a vacuum region more than 36-{\AA}-thick to decouple the neighboring slabs. A $\Gamma$-centered $11 \times 17 \times 1$ $k$-point mesh was used to sample the Brillouin zone. The climbing image nudged elastic band (CI-NEB) method \cite{Henkelman2000} with 5 to 7 intermediate images was used to determine the activation barrier of the phase transition. Besides converging the electronic ground states with charge neutrality, we also considered higher temperature smearing effects and charge doping. More information can be found in Supplementary Note~6 and Figs.~S15--S18. 

In these barrier calculations, the in-plane lattice constants were all constrained to the value of the neutral state 2$H$ phase, even for the charged state conditions in which the lattice tends to expand. We note that the treatment of constant area was mainly due to the difficulty of convergence when the lattice vectors were allowed to relax during CI-NEB calculations of charged slabs. It simulates the situation where the lattice of monolayer MoTe$_2$ is strongly constrained by its interaction with substrates.

Based on the converged electronic ground state with spin-orbit coupling in a given crystal structure, we further constructed the Wannier electronic model using the Wannier90 code \cite{mlwf,mlwf_new} by projecting the Mo $d$ orbitals and Te $p$ orbitals near the Fermi level. Besides capturing the electronic band structure, the Wannier models also allowed us to investigate the topological properties by evaluating the Z$_2$ topological index from the non-Abelian Berry connections along the Wilson loop \cite{Z2_Wilson} and the Fu-Kane parity formulation \cite{FuKane_parity} (when inversion symmetry is present). More information can be found in Figs.~S19--S21.

\section{Acknowledgements}
The authors acknowledge discussions and technical supports from Z.F. Ren, X. Wang, Y. Yoon, B. Pein, P.-C. Shen, A. Maznev, T. Mahony, F. Gao and Y. Chen. J.S. and K.A.N. acknowledge support from the U.S. Department of Energy, Office of Basic Energy Sciences, under Award No.~DE-SC0019126. P.J.-H. acknowledges support from Gordon and Betty Moore Foundation’s EPiQS Initiative through grant GBMF4693. A.Z. acknowledges support from the Miller Institute for Basic Research in Science. W.C., S.F., and E.K. acknowledge support by ARO MURI award W911NF-14-0247. S.F. is also supported by a Rutgers Center for Material Theory Distinguished Postdoctoral Fellowship. J.-C.H. and V.B. acknowledge support from the Center for Energy Efficient Electronics Science (NSF Award 0939514). Y.-Q.B. and Z.C.L. acknowledges support from National Natural Science Foundation of China (No.'s~61974167 and 91963205), the National Key R\&D Program of China (No. 2019YFA0210203) and No.~2019QN01X113. Y.-Q.B. further acknowledges the Open Project of Guangdong Province Key Lab of Display Material and Technology (No.~2020B1212060030). K.W. and T.T. acknowledge support from the Elemental Strategy Initiative conducted by the MEXT, Japan (Grant No.~JPMXP0112101001) and JSPS KAKENHI (Grant No.'s~19H05790 and JP20H00354).

\section{Author Contributions}
Y.-Q.B. and J.S. conceived the idea, conducted the experiments, and analyzed the data under the guidance from P.J.-H. and K.A.N. W.C., S.F., and E.K. provided theoretical calculations. Y.-Q.B., J.H., and V.B. synthesized field enhancement structures. A.Z. and Y.Z. contributed to sample preparations and electron diffraction measurements. Z.C. performed the field enhancement calculations. T.T. and K.W. grew the crystals of hexagonal boron nitride. A.Z., Y.-Q.B., J.S., and E.B. led the manuscript preparation with input from all authors.

\section{Additional information}
Correspondence and requests for materials should be addressed to K.A.N., P.J.-H., and Y.-Q.B. The authors declare no competing financial interests.

\end{document}